\documentclass[superscriptaddress,twocolumn,aps,showpacs,prl,preprintnumbers]{revtex4-1}
\usepackage{epsfig}
\usepackage{graphicx}
\usepackage{amsmath}

\usepackage{xcolor}     
\usepackage{soul}       

\begin{document}

\title{Observation of a disordered bosonic insulator from weak to strong interactions}
\author{Chiara D'Errico}\thanks{These two authors contributed equally.}
\affiliation{LENS and Dipartimento di Fisica e Astronomia, Universit\'a di Firenze, 50019 Sesto Fiorentino, Italy}
\affiliation{Istituto Nazionale di Ottica, CNR, 50019 Sesto Fiorentino, Italy}
\author{Eleonora Lucioni}\thanks{These two authors contributed equally.}
\affiliation{LENS and Dipartimento di Fisica e Astronomia, Universit\'a di Firenze, 50019 Sesto Fiorentino, Italy}
\affiliation{Istituto Nazionale di Ottica, CNR, 50019 Sesto Fiorentino, Italy}
\author{Luca Tanzi}
\affiliation{LENS and Dipartimento di Fisica e Astronomia, Universit\'a di Firenze, 50019 Sesto Fiorentino, Italy}
\author{Lorenzo Gori}
\affiliation{LENS and Dipartimento di Fisica e Astronomia, Universit\'a di Firenze, 50019 Sesto Fiorentino, Italy}
\author{Guillaume Roux}
\affiliation{LPTMS, Univ. Paris-Sud, CNRS, F-91405 Orsay, France}
\author{Ian P. McCulloch}
\affiliation{Centre for Engineered Quantum Systems, University of Queensland, Brisbane 4072, Australia}
\author{Thierry Giamarchi}
\affiliation{DPMC-MaNEP, University of Geneva, 1211 Geneva, Switzerland}
\author{Massimo Inguscio}
\affiliation{LENS and Dipartimento di Fisica e Astronomia, Universit\'a di Firenze, 50019 Sesto Fiorentino, Italy}
\affiliation{INRIM, 10135, Torino, Italy}
\author{Giovanni Modugno}\thanks{modugno@lens.unifi.it}
\affiliation{LENS and Dipartimento di Fisica e Astronomia, Universit\'a di Firenze, 50019 Sesto Fiorentino, Italy}
\affiliation{Istituto Nazionale di Ottica, CNR, 50019 Sesto Fiorentino, Italy}
\date{\today}

\begin{abstract}
We employ ultracold atoms with controllable disorder and interaction to study the paradigmatic problem of disordered bosons in the full disorder-interaction plane. Combining measurements of coherence, transport and excitation spectra, we get evidence of an insulating regime extending from weak to strong interaction and surrounding a superfluid-like regime, in general agreement with the theory. For strong interaction, we reveal the presence of a strongly-correlated Bose glass coexisting with a Mott insulator.
\end{abstract}


\maketitle
The interplay of disorder and interaction in quantum matter is an open problem in physics. A paradigmatic system explored in theory are bosons at $T$=0 in one \cite{Giamarchi88} or in higher dimensions \cite{Fisher}. While disorder alone leads to the celebrated Anderson localization \cite{Anderson}, a weak repulsive interaction can compete with disorder and progressively establish coherence, leading eventually to the formation of a superfluid. A stronger interaction brings however the superfluid into a strongly-correlated regime where disorder and interaction cooperate, leading to a new insulator. The overall phase diagram is therefore predicted to consist in a superfluid, surrounded by a reentrant insulator \cite{Giamarchi88}. The two insulating regimes appearing at weak and strong interaction have been named Bose glass, due to the gapless nature of their excitations. There is an ongoing effort to establish whether they are two distinguishable quantum phases \cite{Cazalilla,Risti,Pollet,Pielawa}. In lattices, the Bose glass at strong interaction is a distinct phase from the gapped Mott insulator appearing at commensurate densities and moderate disorder \cite{Fisher}.

An experimental observation of the reentrant insulator is still missing. Photonic systems and ultracold atoms with tunable weak nonlinearities have demonstrated the coherence effect of a weak interaction on Anderson insulators \cite{Lahini,Deissler}. Magnetic systems with tunable density have provided evidence of the transition from a Mott insulator to a strongly-correlated Bose glass \cite{Hong,Yamada,Yu}, but they lack the possibility to control the interaction. An insulating regime at strong interaction and strong disorder has been detected also with ultracold atoms in tunable optical lattices \cite{Fallani,Pasienski,Gadway}. However, it was so far not possible to distinguish the Bose glass from the Mott insulator.

In this work we study the full problem of disordered interacting bosons in a one-dimensional lattice. We employ ultracold atoms with independently tunable disorder and interaction, which allow us to study systematically the whole disorder-interaction plane. We study several experimental observables and we make a close comparison with the theory. In particular, by means of coherence and transport measurements we identify an insulating regime extending from weak to strong interactions and surrounding a superfluid-like regime. Through a lattice modulation spectroscopy we observe a different interplay of disorder with weak or strong interactions, and in the latter regime we reveal spectral features that are consistent with a strongly-correlated Bose glass coexisting with a Mott insulator. The comparison with theory indicates that the strongly correlated regime is only weakly affected by the finite temperature in the experiment.

We employ an array of quasi-1D samples of $^{39}$K atoms subjected to a quasiperiodic optical lattice \cite{Fallani,Roati08}, which provide a realization of the interacting Aubry-Andr\'e model \cite{Aubry,Modugno}. To a good approximation \cite{Roux,Roscilde}, the dynamics is controlled by the Hubbard Hamiltonian $H=-J \sum_i(b_i^{\dag}b_{i+1}+h.c.)+\Delta \sum_i \cos(2 \pi \beta i)n_i+U/2 \sum_i n_i (n_i-1)+\alpha/2 \sum_i(i-i_0)^2 n_i$, which is characterized by three energy scales: the tunneling energy $J$, the quasi-disorder strength $\Delta$ and the interaction energy $U$. A primary lattice with lattice constant $d=\lambda_1$/2=0.532$\mu$m fixes $J$ ($J/h \simeq 110$Hz). $\Delta$ is essentially the depth of a secondary lattice with an incommensurate wavelength $\lambda_2$ ($\beta$=$\lambda_1/\lambda_2$=1.243). $U$ can be varied from about zero to large positive values thanks to a Feshbach resonance \cite{Roati07}. The fourth term represents a harmonic potential, while $b_i^{\dag}$, $b_i$ and $n_i$ are the creation, annihilation and number operators at site $i$. For $U$=0 all eigenstates are localized above a critical disorder strength $\Delta$=2$J$ \cite{Roati08,Aubry}. Accurate phase diagrams for $U>0$ were obtained theoretically for homogeneous systems and $T$=0 \cite{Roth,Roscilde,Deng,Roux}. The unavoidable trapping in the experiment changes however the nature of the problem, transforming the quantum phases transitions into crossovers and leading to a coexistence of different phases. The 1D systems are populated from an initially three-dimensional Bose-Einstein condensate, which is split into several quasi-1D tubes by an additional 2D lattice.  About 500 such systems are initially created at $U\simeq40J$, then both $U$ and $\Delta$ are slowly changed using almost isoentropic transformations \cite{suppl}. The mean site occupation, $n$=2-7, depends on $U$.

A first indication of the evolution of the system comes from a measurement of the momentum distribution $P(k)$, achieved through absorption imaging after a free flight. The root-mean-square width $\Gamma$ of $P(k)$ is a measure of the coherence of the system. In the superfluid regime (SF) of moderate $U$ and no disorder, comparing $\Gamma$ to calculations \cite{suppl} allows us to estimate the system temperature $k_BT\simeq3J$, which is below the degeneracy temperature for 1D quasi-condensates $k_BT_D\simeq8J$ \cite{Petrov}. Measurements of temperature or even of the presence of thermal equilibrium are instead not possible in the other regimes. The evolution of $\Gamma$ in the disorder-interaction plane in Fig.1(a) shows clearly a coherent regime (blue) for small $\Delta$ and moderate $U$ and an incoherent regime with a reentrant shape from weak to strong $U$ (orange), separated by broad crossovers.

The insulating nature of the incoherent regions is confirmed by transport measurements. These are performed by applying a sudden shift to the harmonic confinement, and detecting the momentum $\delta p$ accumulated in a fixed time interval of 0.9ms, which is essentially the mobility. Figure 1(b) shows $\delta p(U)$ for three different values of $\Delta$.  In the clean case ($\Delta$=0) the motion is almost ballistic for small $U$, while a substantial reduction of the mobility occurs for $U>2-3J$, indicating the progressive formation of an insulator. For finite disorder the mobility at small $U$ is strongly reduced; for increasing $U$ it however increases almost to the clean curve and finally decreases again. This behavior confirms the coherence measurement showing the presence of a disorder-driven insulator at small $U$ and of another insulating regime at large $U$ dominated by the interaction. An additional measurement at a larger $T$, also shown in Fig.1(b), indicates that the mobility for intermediate $\Delta$ is essentially $T$-independent in the accessible range of temperatures.

\begin{figure}[ht!]
\includegraphics[width=0.95\columnwidth] {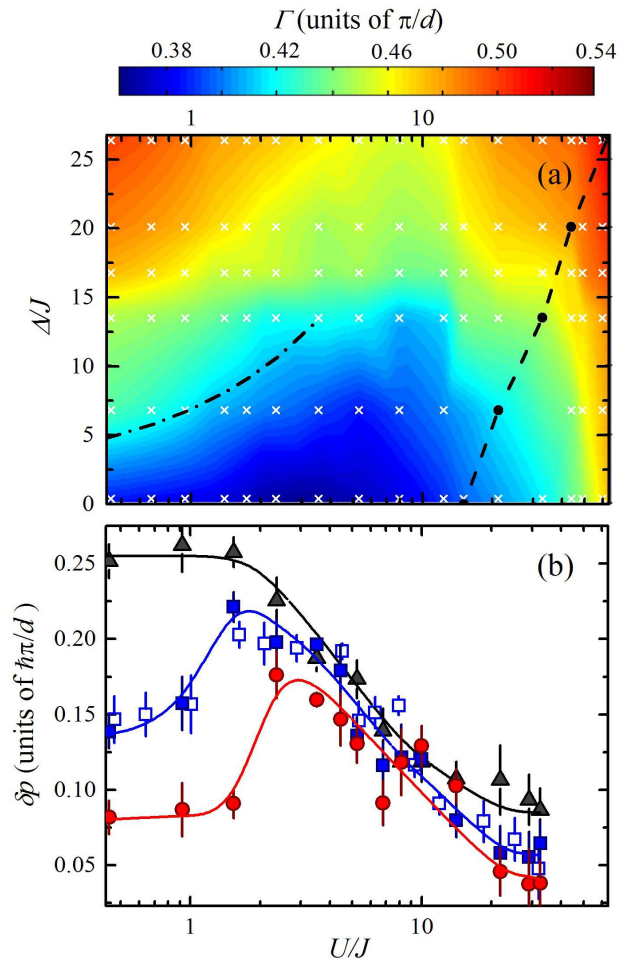}
\caption{Coherence and mobility. (a) Measured width of the momentum distribution. The diagram is built with 94 data points (crosses), with a standard deviation between 2\% and 5\%. $T$=0 calculations reveal a MI only on the right of the dashed line. The dot-dashed line is calculated as $\Delta-2J=nU$. (b) Momentum acquired after an applied impulse for $\Delta$=0 (triangles), $\Delta$=6.2$J$ (squares) and $\Delta$=8.8$J$ (circles), for a SF temperature $k_BT$=3.1(4)J or $k_BT$=4.5(7)$J$ (empty squares). The lines are a guide to the eye. The uncertainties are the standard deviation of typically 5 measurements.}
\label{fig1}
\end{figure}

The overall shape of the incoherent regime in Fig.1(a) is reminiscent of the Bose glass (BG) found in theory at $T$=0 for homogeneous systems \cite{Giamarchi88,Fisher}. The prediction is indeed of a weakly-interacting BG appearing for vanishing $U$ and $\Delta>2J$, which is turned into a SF when the interaction energy $nU$ becomes comparable to the disorder strength, although there is not yet consensus on the exact shape of the phase transition \cite{Lugan,Fontanesi,Vosk}. A stronger interaction is instead expected to lead to a new insulating regime approximately when $U>2nJ$, i.e. when the interaction energy becomes larger than the kinetic energy available in the lattice band. Here, theory predicts a strongly-correlated BG for an incommensurate density (non-integer $n$) and a Mott insulator (MI) for a commensurate density (integer $n$). The latter can however survive in the disorder only up to approximately 2$\Delta<U$, since $U$ controls the MI gap, and the MI disappears for increasing $\Delta$, as shown in Fig.1(a). The opposite trend of the insulating region at large $U$ therefore suggests the appearance of a BG regime in addition to a MI.

\begin{figure*}[t!]
\includegraphics[width=1.9\columnwidth]{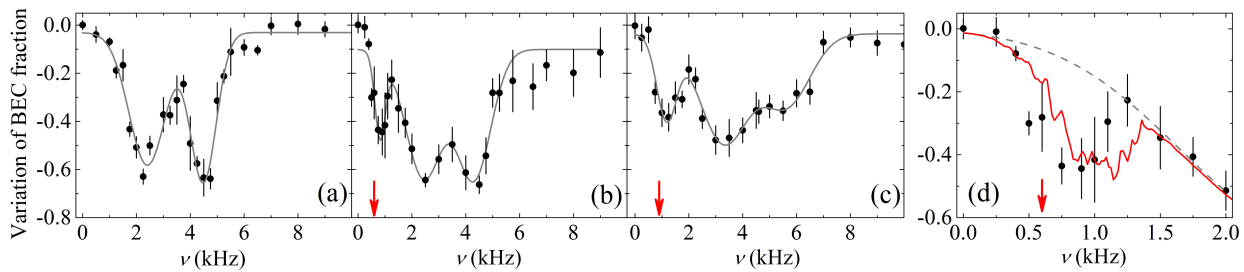}
\caption{(color online) Excitation spectrum for strong interactions. Experimental spectra for $U$=26J and $\Delta$=0 (a), $\Delta$=6.5$J$ (b) and $\Delta$=9.5$J$ (c). The arrows are at $h\nu$=$\Delta$ and the lines are fits with multiple Gaussians. (b) Comparison of the low-frequency peak for $\Delta$=6.5$J$ with theory (continuous line). The theory includes the gaussian tail of the Mott peak (dashed line).}
 \label{fig2}
\end{figure*}

To probe the nature of the insulating regimes, we perform a lattice modulation spectroscopy \cite{Stoferle}. This consists in measuring the energy absorbed by the system when the amplitude of the main lattice, and therefore $J$, is modulated with a sinusoid of variable frequency $\nu$. We start the discussion from the less intuitive large-$U$ regime, summarized in Fig.2. Here the absorption is measured as a decrease of the condensed fraction once the system is transferred back into a 3D trap; we show three characteristic spectra for $U$=26J and increasing $\Delta$. In the clean case one notices the standard MI spectrum \cite{Stoferle,Kollath} with a first excitation peak centered at $h\nu\simeq U$  that can be attributed to MI plateaus with $n$=1-3. A second peak centered at $h\nu\simeq2U$ indicates the coexistence of MI domains with different occupations. The MI domains are connected by incommensurate SF components, which show little response for $h\nu<U$ \cite{Kollath}.

For finite disorder the spectrum changes radically. First, we observe a broadening of the MI peaks by approximately $\Delta$ that indicates an inhomogeneous broadening of the Mott gap, as already observed in previous experiments at strong disorder \cite{Fallani,Guarrera,Gadway}. Second, we observe a striking extra peak appearing in the MI gap, around $\Delta$. This new observation cannot be explained in terms of MI physics, but agrees instead with the expected behavior of a strongly-correlated BG, which can be seen as a weakly-interacting fermionic insulator with a response at about the characteristic disorder energy $\Delta$ \cite{Giamarchi88}. As shown in Fig.2(d), the peak shape is indeed in good agreement with the excitation spectrum of the BG calculated with a fermionized-boson model \cite{Orso,Pupillo}, using the same parameters of the experiment and the temperature measured for the SF. There is a systematic shift of the theory curve to larger frequencies, which might be due to the $U=\infty$ assumption in the model. Within its finite sensitivity, the experiment is compatible with the smooth decrease of the response towards zero frequency associated to the gapless nature of the BG as predicted by the theory.
In the range where we can detect the extra peak, we observe that its center shifts linearly with $\Delta$, as expected in the fermionic picture \cite{Orso}. The peak is no longer detectable for $U/J<$25 as it overlaps with the one at $U$; for very large $U$ the Mott plateaus are instead extending to most of the system, thus lowering the incommensurate fraction and the weight of the peak at $h\nu\simeq\Delta$ \cite{suppl}. In the strong disorder limit, $2\Delta>U$, we observe instead a very broad and essentially featureless spectral response, in agreement with previous experiments \cite{Fallani,Guarrera,Gadway}.

Such response at large $U$ contrasts with the one at small $U$, as shown in Fig.3. Here the energy absorption is measured as an increase of the thermal width of the system. For vanishing $U$, where the system is still globally insulating, we observe a weak excitation peak centered at $\Delta$, which is however already broader than the one predicted for non-interacting bosons \cite{suppl}, especially towards small $\nu$. This suggests the formation of large coherent regions due to the coupling of single-particle states by the interaction, leading to the possibility of long-distance, small-$\nu$ excitations. This behavior recalls the prediction for the $T$=0 weakly-interacting BG being composed by large, disconnected SF regions. A moderate increase of $U<\Delta$ leads indeed to a rapid broadening of the response, with a strong enhancement for small $\nu$. This behavior is in clear contrast with that of the strongly-interacting insulator, where the weak response at small-$\nu$ indicates a strong fragmentation. A further increase of $U$ eventually brings the system into a regime spectrally indistinguishable from a clean SF, confirming the delocalizing role of the interactions in this regime \cite{Deissler,Tanzi}.

\begin{figure}[ht!]
\includegraphics[width=0.9\columnwidth]{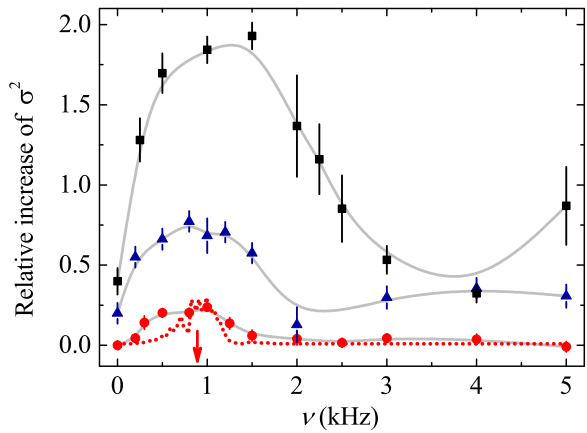}
\caption{(color online) Excitation spectrum for weak interactions. Spectra for $\Delta$=7$J$, and $U$=0.35$J$ (circles), $U$=1.4$J$ (triangles), or $U$=2.1$J$ (squares). The dataset have been shifted vertically by 0.2 for clarity. The dotted red line is the theory for non-interacting bosons, the arrow is at $h\nu$=$\Delta$ and the continuous grey lines are a guide to the eye. }
 \label{fig3}
\end{figure}

The temperature cannot be measured for finite $\Delta$, but we can still gain an insight in thermal effects by comparing the experimental $P(k)$ to the exact $T$=0 theory supplemented by a phenomenological account of a finite temperature. To do this, we apply a density-matrix renormalization group technique (DMRG) \cite{Schollwock,McCulloch,Roux} to simulate an ensemble of trapped systems with the same parameters of the experiment. Figure 4(a-d) show a few representative comparisons of the experimental $P(k)$ and the theoretical ones at $T$=0. In the low-$U$ region, the calculated $P(k)$ is definitely narrower than the measured one, whereas in the strongly correlated region the broadening is less relevant. Assuming thermal equilibrium, we quantify this thermal broadening by convolving the calculated $P(k)$ with a Lorentzian distribution corresponding to an exponential decay of the correlations with a thermal length $\xi_T$, an approach known to be valid for the SF \cite{Giamarchibook}. The dashed red lines in Fig.4(a-d) are the best fit of the theory to the experiment with $\xi_T$ as the only fitting parameter. There is a good agreement in all regimes, except for the one at small $\Delta$ and large $U$ (Fig.4(c)). For small $U$, the thermal length is short ($\xi_T\simeq d$), revealing a relevant thermal excitation of both SF and insulating regimes. In the large-$U$ region, $\xi_T$ is instead large, suggesting that the strongly-correlated phases are only weakly affected by the finite $T$.

\begin{figure}[ht!]
\includegraphics[width=0.9\columnwidth] {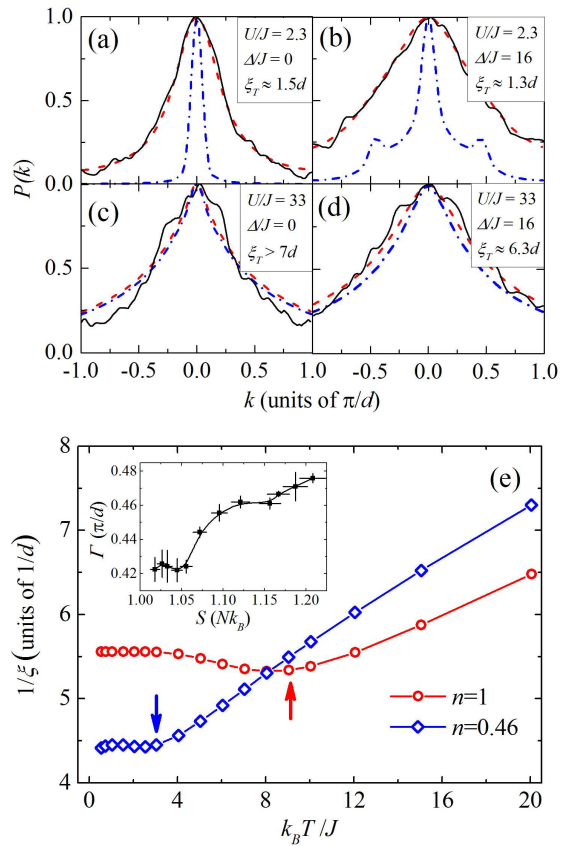}
\caption{ (color online) Finite-temperature effects. (a-d) Comparison of the experimental momentum distributions (black, continuous) with $T$=0 DMRG calculations (blue, dash-dotted) supplemented by the introduction of a thermal correlation length $\xi_T$ (red, dashed), for four points in the $\Delta-U$ plane. (e) Temperature evolution of the correlation length calculated by exact diagonalization for $U$=44J and $\Delta$=10$J$ for two different homogeneous densities; arrows: crossover temperatures $T_0$. Inset: measured width of $P(k)$ for $U$=23.4J and $\Delta$=6.6$J$ vs entropy per particle. The corresponding SF temperature varies from 3.1(1)$J$ to 4.7(2)$J$. The line is a guide to the eye.}
 \label{fig4}
\end{figure}

An exact diagonalization of the Hamiltonian in Eq.1 for small homogeneous systems confirms the latter indication \cite{suppl}. Fig.4(e) shows that for large $U$ and $\Delta$ the correlation length $\xi$ is only weakly dependent on $T$ at low temperatures and a relevant broadening appears only above a crossover temperature $T_0$. This can be clearly seen not only for the MI, where a simple explanation in terms of the energy gap exists \cite{Gerbier}, but also for the gapless BG. This apparently surprising result is justified by the fact that the effective Fermi energy of the BG is of the order of $\Delta$, and therefore it should not be too sensitive to $k_BT<\Delta$. A detailed study we performed shows that $T_0$ for the BG scales indeed linearly with $\Delta$; for small $\Delta$ it becomes negligible, justifying why the single-$\xi_T$ approach does not work in the regime of Fig.4c. In the experiment, a study of $\Gamma$ for the coexisting BG and disordered MI as a function of the entropy (Fig.4(e)) shows the existence of a plateau at low entropy, before a broadening sets in. These results suggest that the $T$=0 quantum phases persist in the experiment for sufficiently large $U$ and $\Delta$.

An analogous study we performed for small $U$ confirms the presence of much larger thermal effects \cite{suppl}. It is however interesting to note that the broadening of $P(k)$ is not accompanied by a change of the mobility (see Fig.1(b)). In the future, it will be interesting to study the possible relation of this persisting insulating behavior at finite $T$ with the proposed many-body localization \cite{Aleiner,Michal}.

In conclusion, we have shown evidences of the reentrant insulator predicted for disordered bosons. The strongly-interacting regime shows excitation properties as predicted by the $T$=0 theory for the Bose glass. It is possible to apply these techniques to further studies of disordered bosons, such as in the absence of a lattice, to study the Bose glass in one dimension without overlap with the Mott physics, or in lattices of higher dimensionality, where the superfluid is expected to be much more resistant to disorder \cite{Fisher,Pollet09}.

This work was supported in part by the ERC (grants 203479 and 247371), by the Italian MIUR (PRIN2009FBKLN and RBFR12NLNA) and by the Swiss NSF under MaNEP and Division II. G.R. was supported by grant ANR-2011-BS04-012-01 QuDec. I.MC. was supported by the ARC Centre of Excellence for Engineered Quantum Systems (grant CE110001013). The experimentalists in Florence acknowledge discussions with S. Chauduri, L. Fallani, C. Fort and M. Modugno, and contributions by A. Kumar. T.G. is grateful to the Harvard Physics Department and the MIT-Harvard Center for Ultracold Atoms for support and hospitality.

\newpage

\setcounter{figure}{0}
\renewcommand\thefigure{S\arabic{figure}}
\section*{Supplementary material}

\textbf{System preparation}. We prepare a Bose-Einstein condensate of $^{39}$K ground-state atoms in a 3D harmonic trap with mean frequency $\omega$=2$\times$80 Hz, at a scattering length $a$=210$a_0$. A strong 2D lattice with spacing $d$=$\lambda_1$/2 is then slowly raised to a typical height of 30 recoil energies. The system gets split into about 500 elongated traps (tubes), with an overall Thomas-Fermi distribution. The estimated atom number distribution is $N_{i,j}=N_{0,0} [1-2\pi N_{0,0} (i^2+j^2)/5N_T ]^{3/2}$, where $i$ and $j$ are the tube indexes, $N_T$ is the total atom number and $N_{0,0}$ is the atom number in the central tube. $N_T$ is in the range (2-4)$\times10^4$, depending on the specific dataset. From a simulation of the lattice loading procedure we estimate an upper limit of $N_{0,0}$=96(20) for $N_T$=3(1)$\times10^4$.
The radial trapping frequency of each tube, $\omega_\perp$=2$\pi\times$50 kHz, is larger than any other energy scale (including the axial trapping frequency, typically $\omega_z$=2$\pi\times$150 Hz), and tunneling between neighboring traps is suppressed on the timescale of the experiments ($h/J_\perp$0.5 s). The mean atomic density for each 1D system is estimated as the largest of the mean field and the Tonks value \cite{Dunjko}. The mean site occupation n is then calculated by averaging over all subsystems.
\begin{figure}[ht!]
\includegraphics[width=0.95\columnwidth] {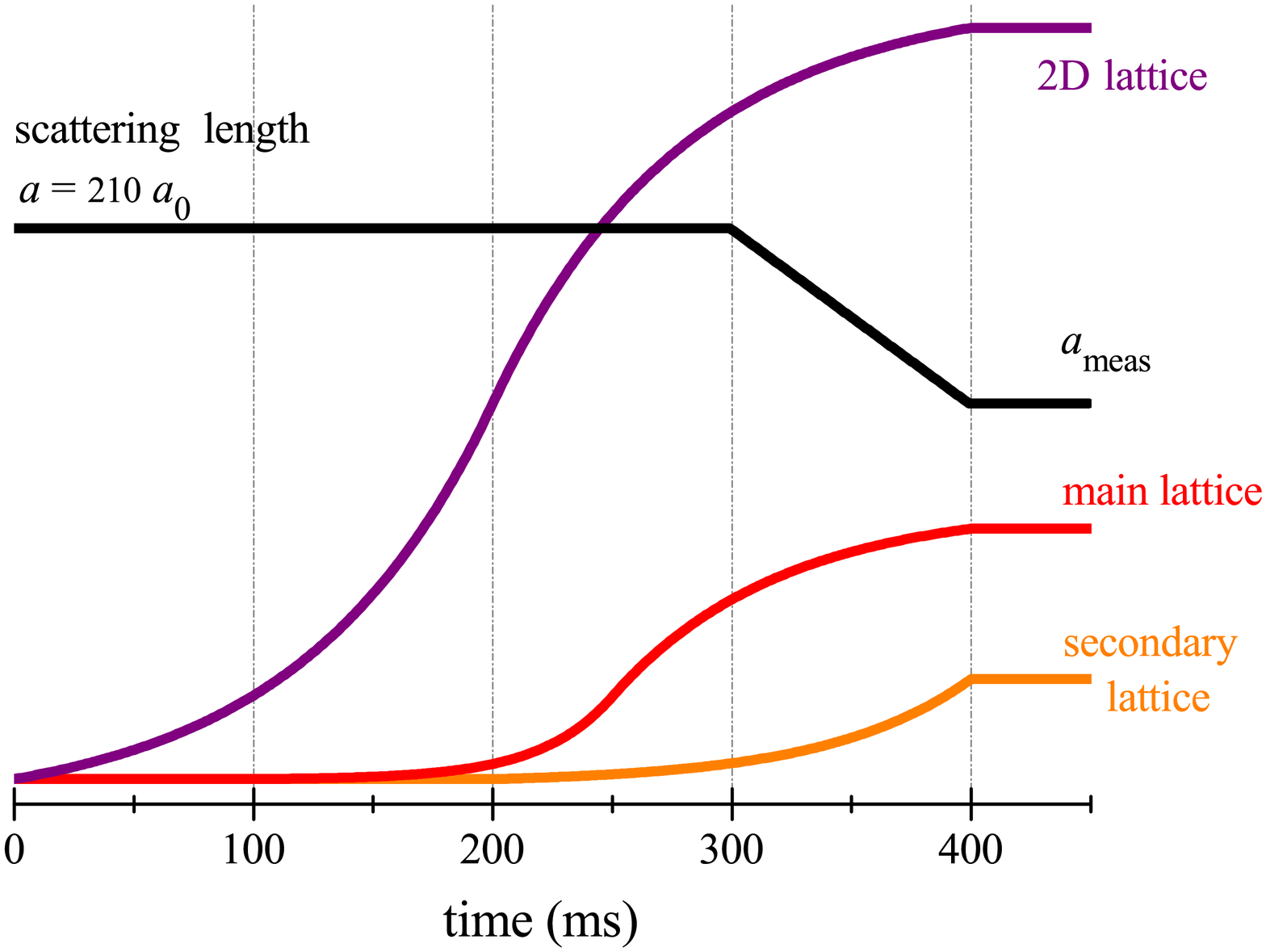}
\caption{Experimental preparation sequence. The loading ramp for the 2D lattice (purple) is s-shaped, with a total duration of 400 ms. Half of the total depth is reached with an exponential ramp lasting 200 ms, with time constant $\tau$=80 ms; the final depth is then reached with an inverted exponential with the same $\tau$. The main lattice starts to raise 100 ms after the 2D lattice. Its ramp (red) is s-shaped, with a total duration of 300 ms. One third of the total depth is reached in 150 ms with an exponential ramp with $\tau$=30 ms; the final depth is reached with an inverted exponential ramp with $\tau$=-70 ms. The secondary lattice (orange) raises during the last 200 ms, with an exponential ramp with $\tau$=60 ms. The scattering length (black) is linearly changed to the desired value after 300 ms, when the radial confinement is sufficiently strong to freeze the number of occupied tubes.}
\label{figS1}
\end{figure}
A quasiperiodic lattice is also slowly applied along the tubes, consisting of a main lattice with spacing $d$=$\lambda_1$/2 and a depth of about 9 recoil energies, which sets the tunneling energy $J$ ($J/h\simeq$110 Hz), and a secondary lattice with $\lambda_2$=$\lambda_1/\beta$=856 nm, whose depth sets the quasi-disorder strength $\Delta$ \cite{Modugno}. The mapping of the first band of such lattice onto the Hamiltonian in the main paper is not perfect, since there is in principle also a spatial modulation of the hoppings between neighboring sites \cite{Roux}. However, the modulation of $J$ we calculate is only 7\% for the largest disorder strength in the experiment, and it is probably negligible at moderate disorder strengths.

The experimental sequence for the lattices preparation is summarized in Fig.\ref{figS1}. All the parameters have been optimized to have the most adiabatic loading \cite{Gericke}. Both $J$ and $\Delta$ are calculated from the measured lattice depths. The 2D lattice is calibrated by means of Raman-Nath diffraction while the main and secondary lattices are calibrated with Bragg oscillations. The typical error on both 2D and main lattice depths is 2\%. The typical error on the secondary lattice depth is 20\%. This results in a typical error of 20\% on $\Delta/J$. We found that the relative phase of the two lattices forming the quasiperiodic potential normally fluctuates in successive measurements. We employ this mechanism to provide averaging over the quasi-disorder. During the preparation, the interaction energy  $U=(\hbar^2/m a_{1D})\int|\phi(x)|^4 dx$ is slowly changed to the desired value by changing the 1D scattering length $a_{1D}=a_\perp^2 (1-1.03 a/a_{\perp})/2a$, via a change of the 3D scattering length $a$ at a Feshbach resonance \cite{Roati07}. Here $\phi(x)$ is the single-particle Wannier function and $a_\perp=\sqrt{\hbar/m\omega_\perp}$. The typical uncertainty on $U/J$ is 6\%.
A characteristic temperature for quantum degeneracy for one-dimensional bosons in the SF regime, can be estimated as $k_B T_D$=(3/16)$\hbar\omega^* N$, where $\omega^*$=$\omega_z/\sqrt{m^*/m}$ is the lattice-renormalized axial frequency \cite{Petrov}. The typical value for the experimental parameters is $T_D$=40(7) nK.

\begin{figure}[t!]
\includegraphics[width=0.95\columnwidth] {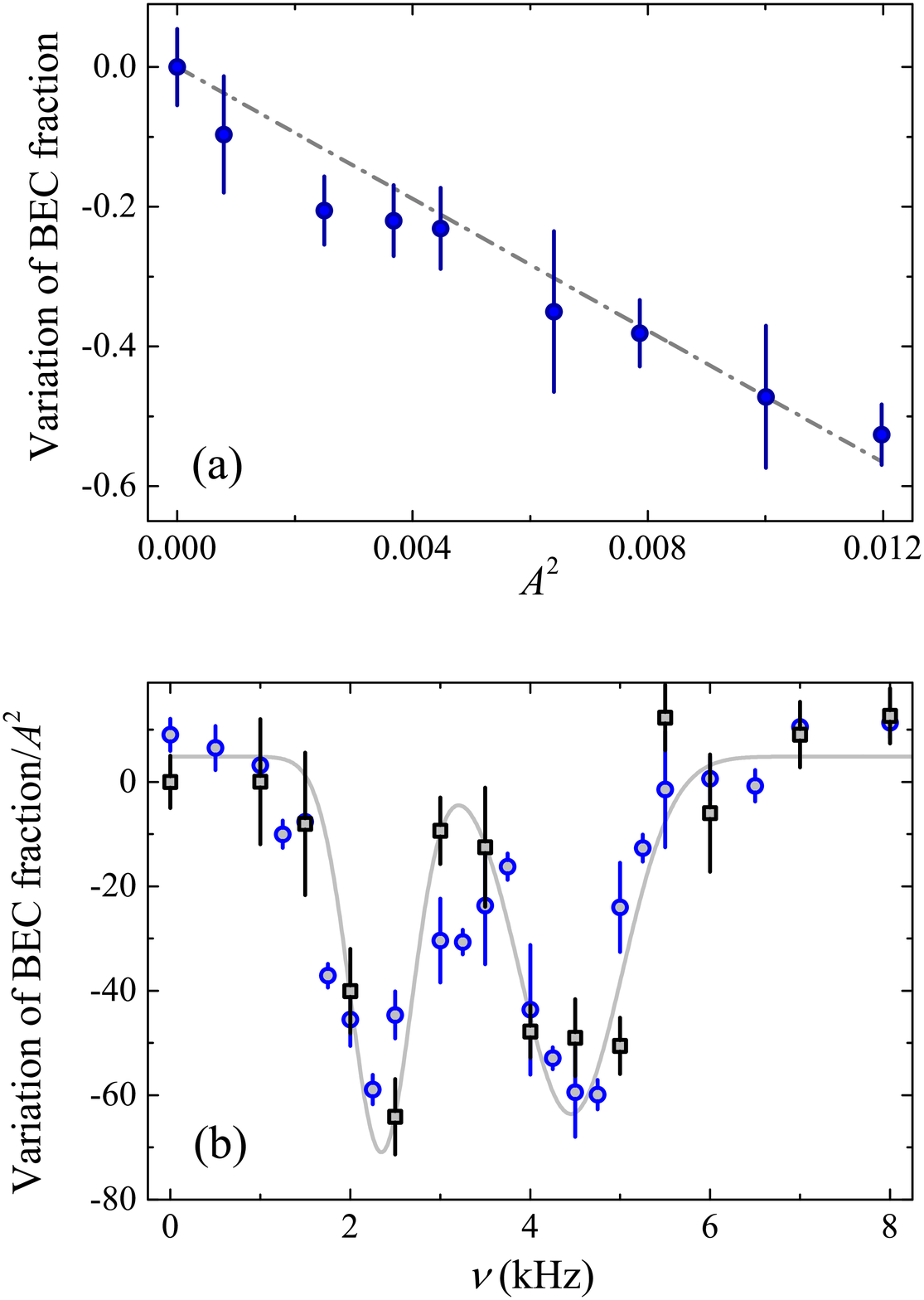}
\caption{Linear response. (a) Relative variation of the condensed fraction for $\Delta$=0 and $U$=5.4$J$, versus the square of the modulation amplitude.  (b) Excitation spectrum for $\Delta$=0 and $U$=26$J$, for two different modulation amplitudes, $A$=0.05 (black) and $A$=0.1 (blue), normalized to the square of the modulation amplitude.}
\label{figS2}
\end{figure}

\textbf{Excitation spectrum}. The excitation spectrum is measured by modulating the depth of the main lattice as  $V(t)$=$V_0 (1+A \cos⁡(ωt))$, with $A\simeq$0.1. We checked that such modulation amplitude is small enough to stay in the linear response regime, see Fig.\ref{figS2}.
In order to have the maximum sensitivity at low frequency, we used the longest modulation time of 200 ms allowed by the background heating of the system.  The absorbed energy is extracted from the temperature increase, which is measured after having retransferred the atoms back to the 3D trap. In the BEC regime, the temperature is estimated from the measured condensed fraction $\eta(\omega)$. Since $\eta<$0.3 we use the first order approximation $\Delta T(\omega)$=-$T_c(\eta(\omega)-\eta(0))$ where $\eta(0)$ is the unperturbed condensed fraction. In the thermal regime of $T/T_c >1$, employed for the measurements at low $U$ in Fig.3, we use the thermal width $\sigma(\omega)$: $\Delta T(\omega)$=$m(\sigma^2(\omega)-\sigma^2(0))/k_Bt_{exp}^2$, where $t_{exp}$ is the expansion time and $\sigma(0)$ is the unperturbed width.
The excitation spectrum of the strongly-correlated Bose glass has been modeled in the limit of non-interacting fermions. We employ the absorption rate derived in linear approximation \cite{Orso}:
$\dot{E}(\omega)=(\delta J^2\pi\omega)/2 \sum_{a,b}\overline{K_{a,b} [f_{FD} (\epsilon_a )-f_{FD} (\epsilon_b )]\delta(\hbar\omega+\epsilon_a-\epsilon_b)}$.
Here $K_{a,b}=|\sum_i(\phi_a^* (i+1) \phi_b (i)+\phi_a^* (i) \phi_b (i+1)|^2$ is calculated over pairs $(a,b)$ of single-particle eigenstates of the quasiperiodic lattice, $f_{FD} (\epsilon)$ is the Fermi-Dirac distribution at finite $T$, and the bar represents averaging over different realizations of the potential. We adapted the same model above to ideal bosons. In this case we replaced $f_{FD} (\epsilon_a )-f_{FD} (\epsilon_b )$ with the Bose-Einstein distribution $f_{BE} (\epsilon_a )$. The spectra in Fig.2(d) and Fig.3 are calculated on a 200-sites lattice, with the same harmonic trap as in the experiment, and $T$ as measured in the SF region.
\begin{figure}[ht!]
\includegraphics[width=0.9\columnwidth] {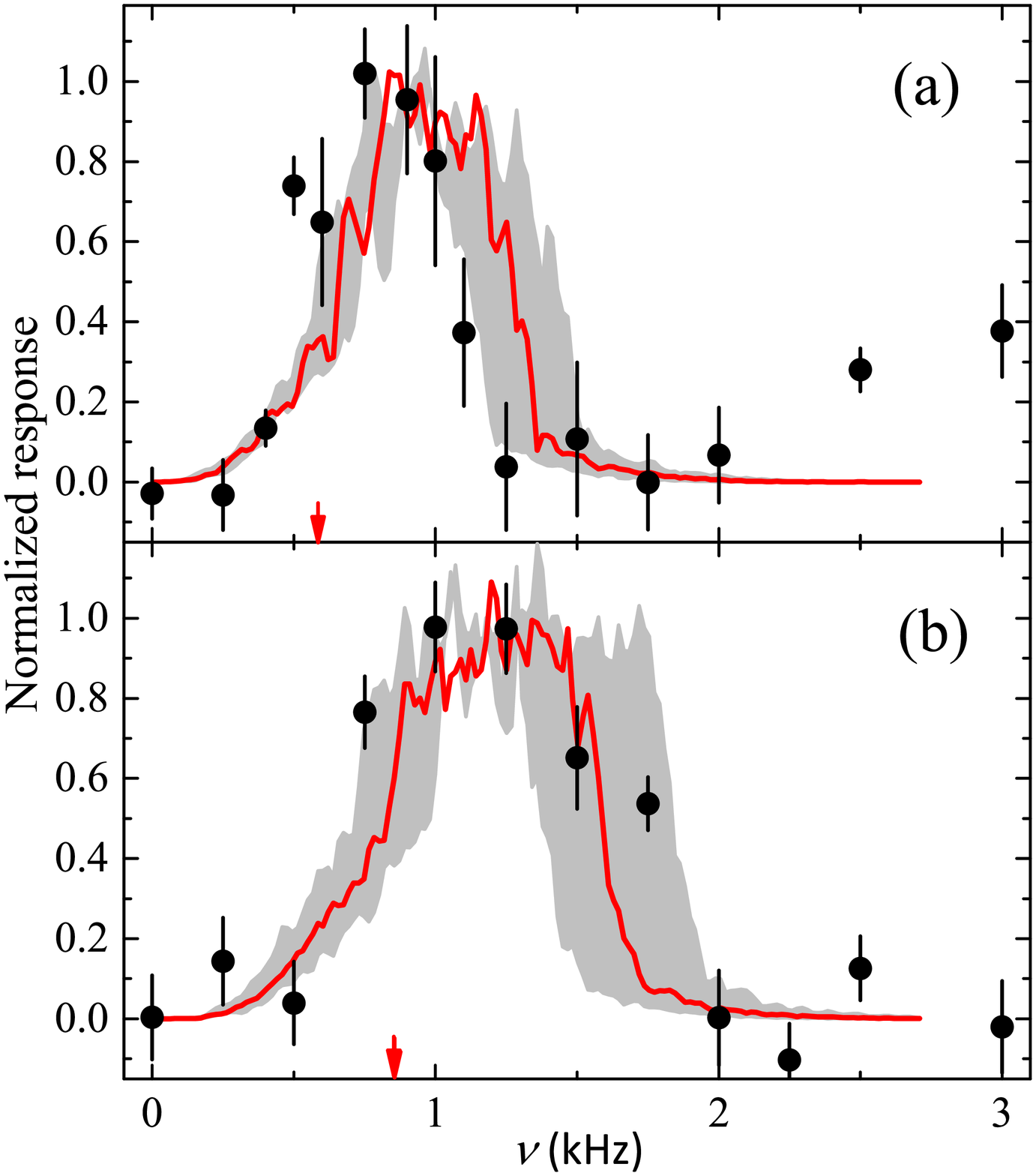}
\caption{Excitation spectrum of the Bose glass. Experiment-theory comparison for the low-frequency part of the spectra in Fig.2, for two disorder strengths: (a) $\Delta$=6.5$J$ and (b) $\Delta$=9.5$J$. The red curve is the spectrum calculated for the nominal $\Delta$, and the grey region shows the effect of the 20\% uncertainty on $\Delta$. The arrows mark the disorder strength in frequency units, $\Delta/h$.}
\label{figS3}
\end{figure}

\begin{figure*}[ht!]
\includegraphics[width=1.9\columnwidth] {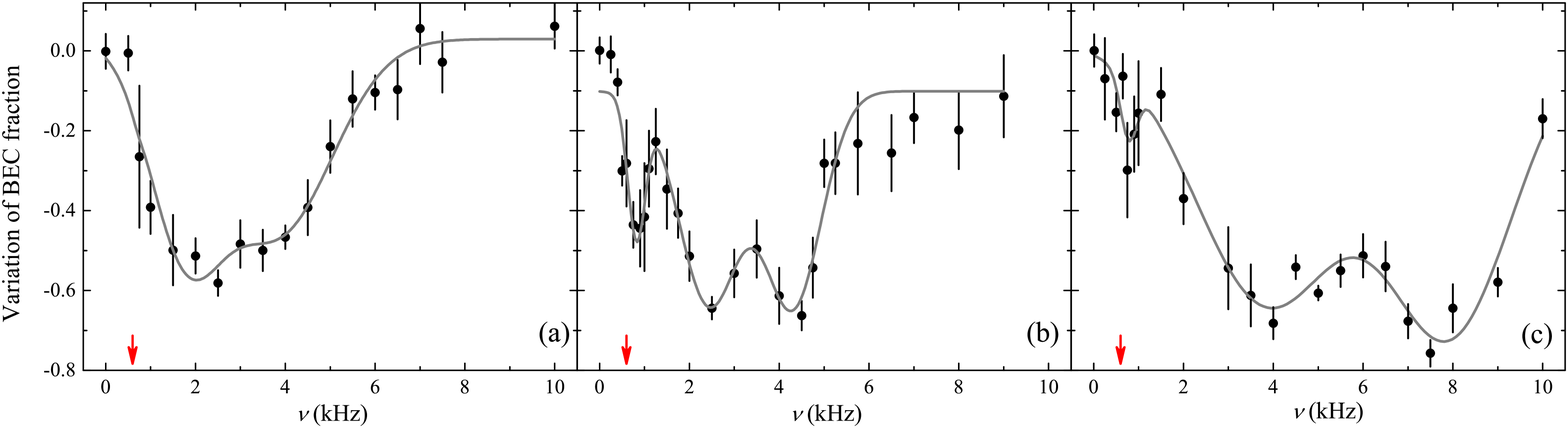}
\caption{Excitation spectrum for varying interaction in the strongly-correlated regime. The excitation spectrum, measured as the relative variation of condensed fraction, for $\Delta$=6.5$J$ and three different interaction energies: (a) $U$=20$J$, (b) $U$=26$J$ and (c) $U$=58$J$. The tunneling energy corresponds to a frequency $J/h$=90 Hz, and the arrows mark $\Delta/h$.}
\label{figS4}
\end{figure*}

In the strongly-correlated regime, the fermionization approach requires an infinite repulsive interaction. To account for the finite $U$ in the experiment, which implies a site occupation $n>$1, we have employed an extended-fermionization approach \cite{Pupillo}. The idea is to neglect the coupling between layers with different occupations ($n\leq1$, $1<n\leq2$, $2<n\leq3$), and to calculate their response independently, taking properly into account the larger kinetic energy of the excited bands of the fermionic model. This approach has been shown to work rather well already for relatively small $U$, provided that the Mott gap is open. The spectrum shown in Fig.2(d) has been first calculated for individual tubes, and then averaged over the distribution of tubes in the experiment. We verified that the density distributions in individual tubes obtained with the extended-fermionization approach are very close to the ones calculated by DMRG. The vertical scale of the simulation has been rescaled to match the experimental observation, and a Gaussian background has been added, to take into account the first Mott peak.

A further comparison of theory and experiments for two different disorder strengths in the strongly correlated regime ($U$=26$J$) is shown in Fig.\ref{figS3}. Here we subtracted from the experimental data the Gaussian background of the MI peak, and we normalized the resulting peak response to unity. There is a clear shift and a broadening of the peak with increasing $\Delta$. The experimental data are slightly shifted to larger frequencies than the theory predictions. This might be due to the finite-$U$ in the experiment. However, the present uncertainties in the experimental parameters do not allow drawing any conclusion in this direction, since the uncertainty on $\Delta$ reflects in an uncertainty on the peak position. A similar effect, although smaller, is also introduced by the total atom number and trap frequency uncertainties.

The evidence of a strongly-correlated BG from the $\Delta$-peak in the excitation spectrum can be gained only in a limited region of $U-\Delta$ values, as shown for example in Fig.\ref{figS4}. Indeed, when $\Delta$ is comparable with $U$, the MI peak overlaps with the BG one (Fig.\ref{figS4}(a)). When instead $U$ is much larger than $\Delta$, the incommensurate density component that can form a BG is strongly reduced (Fig.\ref{figS4}(c)). This limited range of observation of the $\Delta$-peak is presumably the reason why it was not observed in previous experiments \cite{Fallani,Gadway,Guarrera}.

\begin{figure}[ht!]
\includegraphics[width=0.95 \columnwidth] {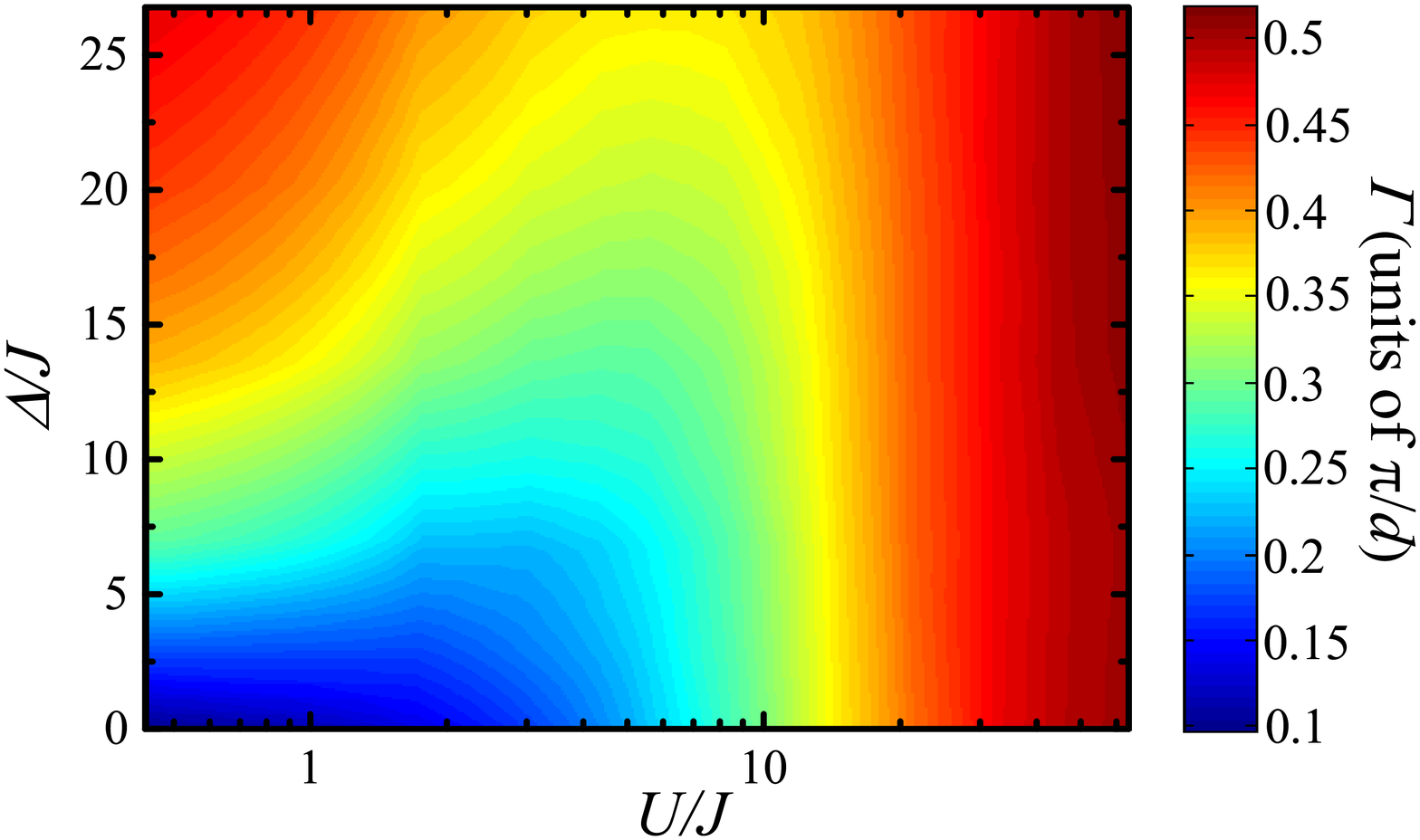}
\caption{Theoretical coherence at $T$=0 from DMRG calculations.  Width $\Gamma$ of $P(k)$ at $T$=0, calculated for individual tubes, and then averaged over the distribution of tubes. The diagram is built with 94 data points at the same positions of the experimental data in Fig.1(a). }
\label{figS5a}
\end{figure}

\textbf{$U-\Delta$ diagram from DMRG calculations}. Zero-temperature DMRG calculations give access to all single-particle correlations $g_{ij}$=$\langle b_i^\dagger b_j\rangle$ in the ground-state and to the density profiles. Also the $T$=0 momentum distribution can be obtained using $P(k)$= $|W(k)|^2\sum_{lm}\exp(ik(l-m)) g_{lm}$, where $W(k)$ is the Fourier transform of the numerically computed Wannier function, and averaging over all tubes. This calculation gives the blue curves in Fig.4(a-d). The change of the rms width $\Gamma$ of the calculated $P(k)$ across the $\Delta-U$ plane is shown in Fig.\ref{figS5a}.

\begin{figure}[ht!]
\includegraphics[width=0.95\columnwidth] {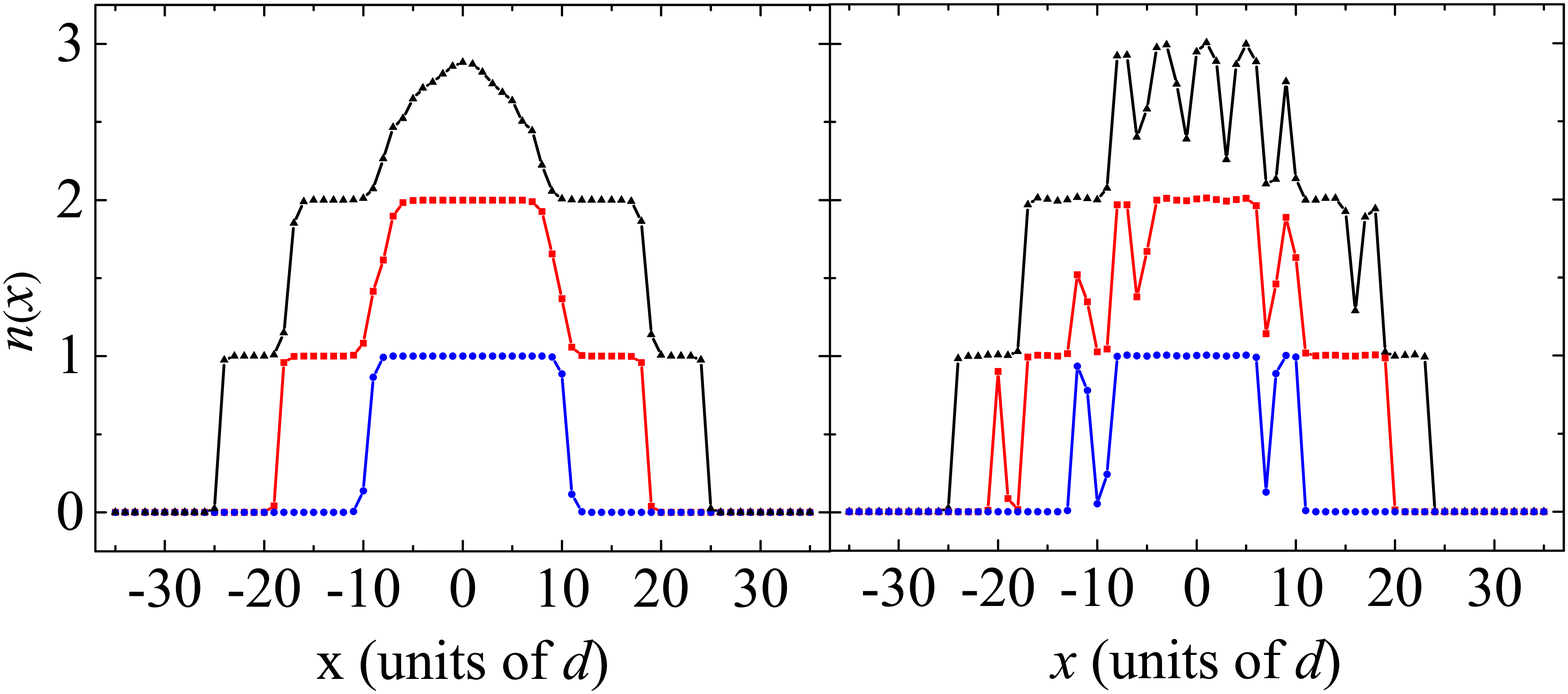}
\caption{Density profiles obtained from DMRG calculations for $U$=26$J$ and $\Delta=0$ (left panel) or $\Delta=6.5 J$ (right panel). Blue, red and black curves refer respectively to $N$=20,55,96.}
\label{figS10}
\end{figure}

In Fig.\ref{figS10} we show the calculated density profiles in tubes with $N$=(20, 55, 96), for the clean case and for $\Delta=6.5 J$. By looking at the density profiles one can note that, adding disorder, the Mott regions progressively shrink and the smooth profiles of the incommensurate regions between the different Mott insulator shell are turned in strongly irregular ones, as expected in the case of a Bose glass. The dashed line in Fig.1(a) is obtained by looking at the presence of Mott insulating regions in the density profiles for these three representative tubes. Mott regions are defined by three consecutive sites with density almost equal to an integer filling. Notice that due to the trap, the value of $U$ required to stabilize Mott regions is larger than the homogeneous result for the average value of 2 atoms per site $U$=3.3$J$. The dash-dotted line in Fig.1(a) is instead calculated as $\Delta-2J$=$nU$, where the offset 2$J$ is put to tentatively take into account the finite disorder strength needed to localize the non-interacting particles \cite{Tanzi}.

\begin{figure}[ht!]
\includegraphics[width=0.95 \columnwidth] {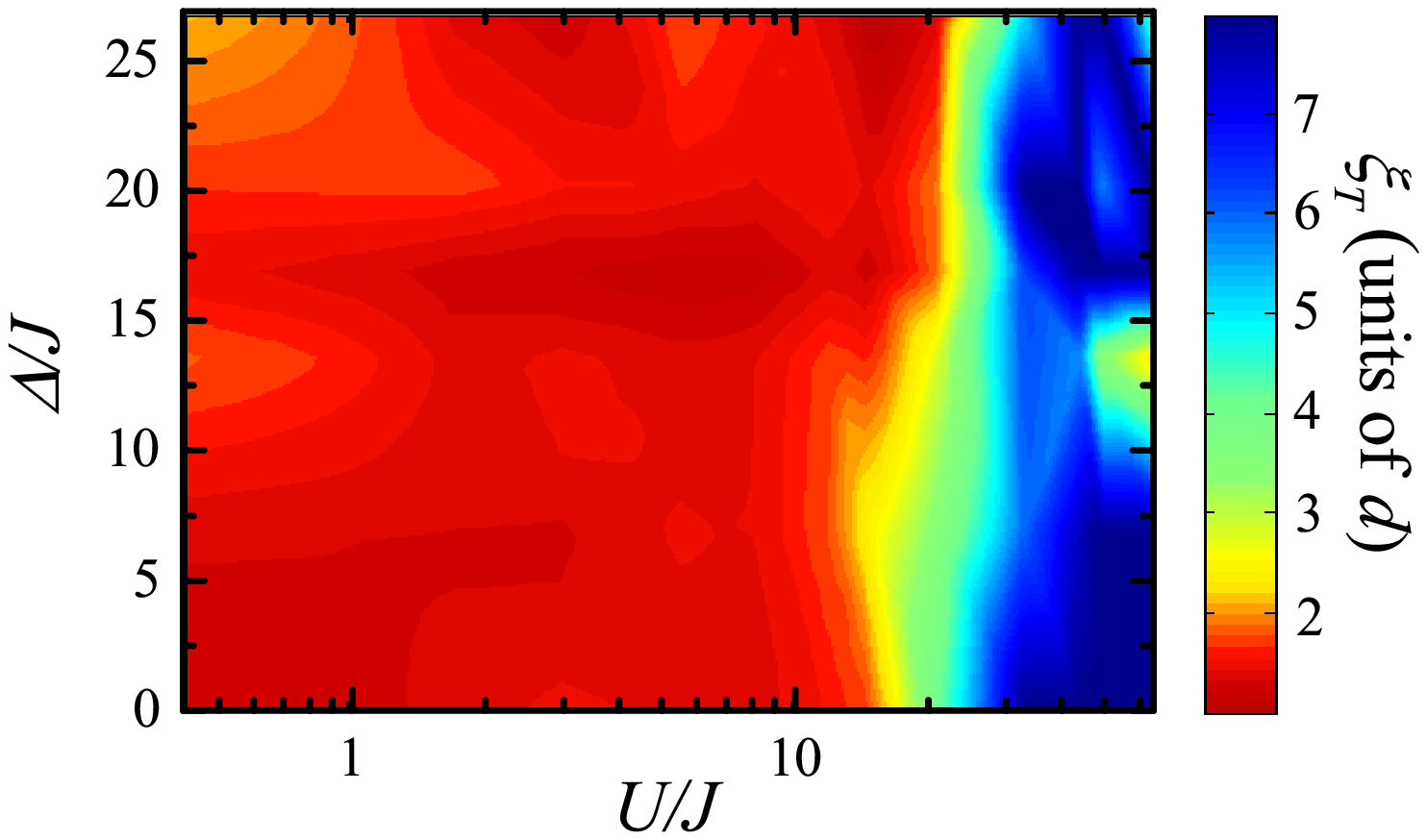}
\caption{Thermal correlation length $\xi_T$ fitted in the experiment-theory comparison. Thermal effects are more relevant for small $U$.}
\label{figS5b}
\end{figure}

A phenomenological account of the temperature effect is introduced by multiplying the $g_{lm}$ above by $\exp(-|l-m|/\xi_T )$ with $\xi_T$ a thermal correlation length. For sake of simplicity, one can take the same  $\xi_T$ for all distances and all tubes. This qualitatively amounts to convolve the $T$=0 momentum distribution with a Lorentzian of width 1/$\xi_T$. We fit the experimental profiles leaving $\xi_T$ as a free fitting parameter (red curves in Fig.4(a-d)). The fitting results across the $\Delta-U$ diagram are shown in Fig.\ref{figS5b}. As discussed in the manuscript one can clearly distinguish two regimes: when $U$ is small $\xi_T$ is short (red region of the diagram) while it increases significantly when $U$ is large (blue region). This indicates the fact that temperature strongly affects the correlation properties in the weakly-interacting regime while only slightly perturbs the strongly-correlated ones. The different impact of the finite temperature in the two regions is confirmed by exact diagonalization.

\begin{figure}[ht!]
\includegraphics[width=0.9\columnwidth] {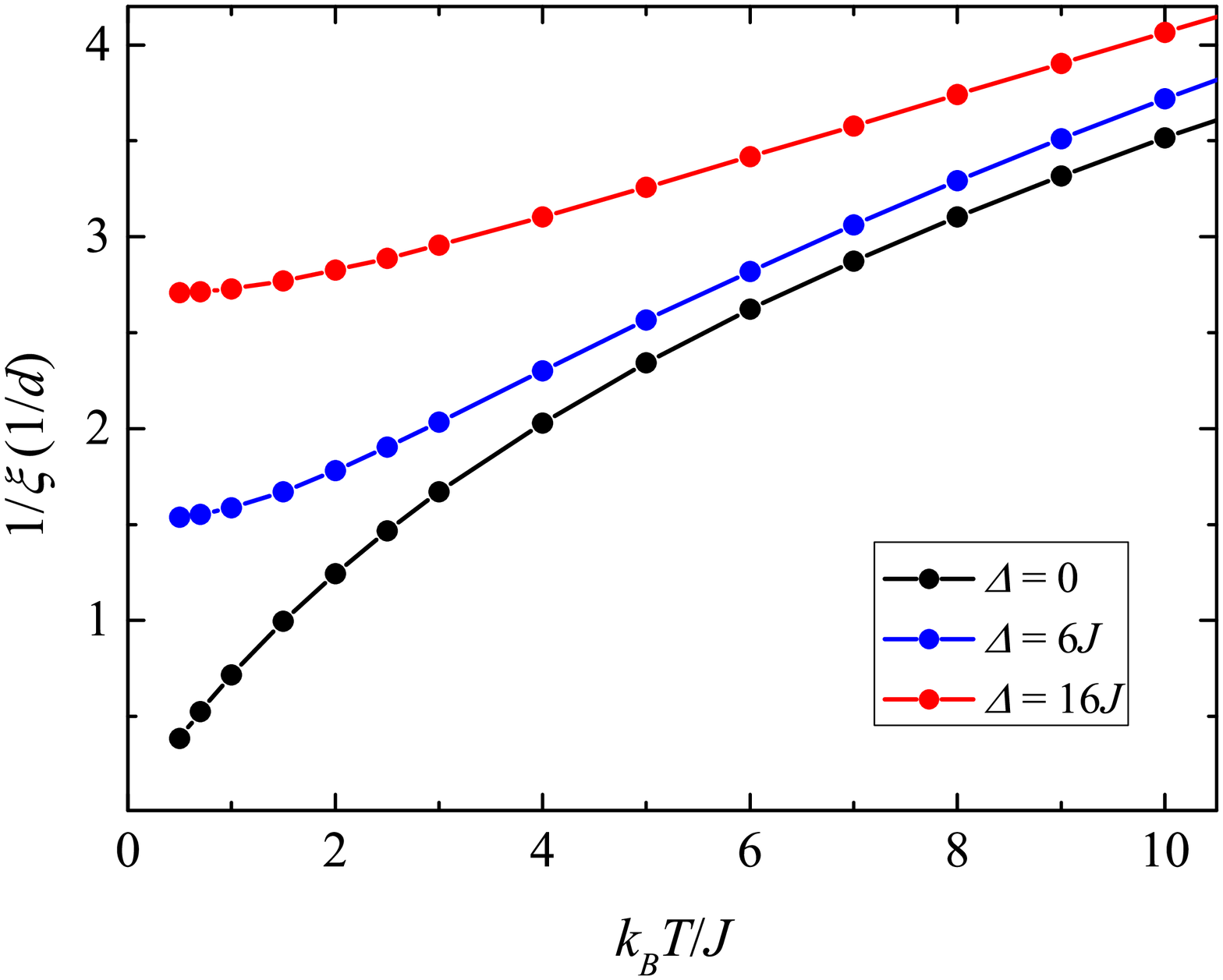}
\caption{Weakly interacting correlation length vs disorder and temperature. Correlation length $\xi$ vs $T$, calculated by exact-diagonalization of a weakly-interacting system with $U$=2.3$J$ and $n$=0.46.}
\label{figS6}
\end{figure}

\textbf{Correlation length from exact diagonalization at finite temperature}. The exact diagonalization study is performed on small homogeneous systems with length up to $L$=12$d$ and mean site occupations up to $n$=1. We perform calculation for one representative weak interaction ($U$=2.3$J$) and one representative strong one ($U$=44$J$). We calculate the temperature evolution of the correlation function for various disorder strengths and we extract the correlation length $\xi$ from an exponential fit of the tails.
In Fig.\ref{figS6} we report representative results for the small $U$ case. The inverse correlation length starts to increase already for very small temperatures, implying a non-negligible impact of thermal fluctuations on the $T$=0 quantum phases and justifying the experimental observation of a rather short $\xi_T$ for weak interactions.

\begin{figure}[ht!]
\includegraphics[width=0.95\columnwidth] {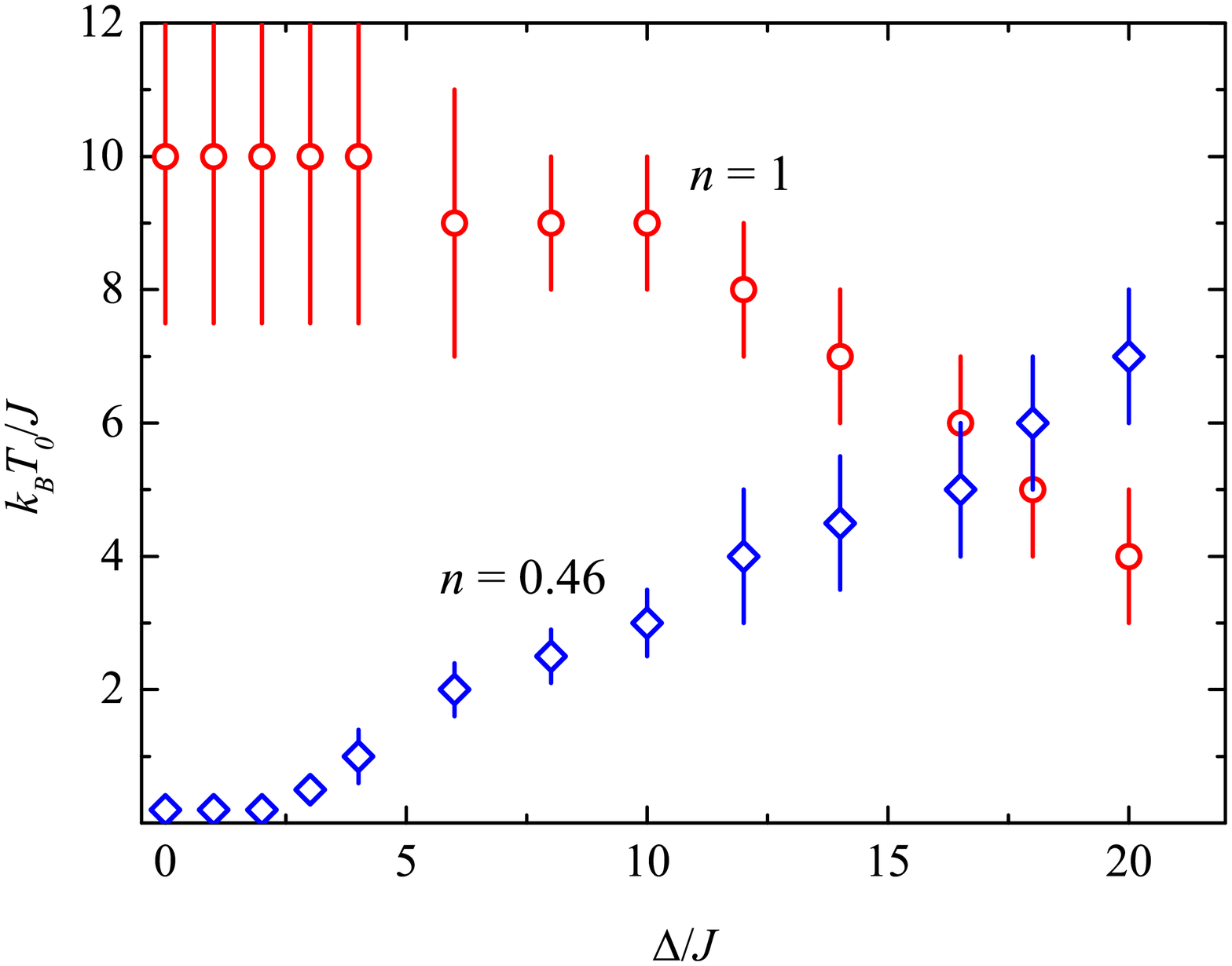}
\caption{Scaling of the crossover temperature with $\Delta$. Crossover temperature $T_0$ vs the disorder strength, calculated by exact-diagonalization of a strongly interacting system ($U$=44$J$), for two commensurate and incommensurate densities.}
\label{figS7}
\end{figure}

On the contrary, in the strongly-interacting case the calculated $\xi$ remains almost constant up to a finite crossover temperature $T_0$, indicating that the thermal effect on the correlation of the system starts to appear only above $T_0$ (Fig.4(e)). We extract $T_0$ as the maximum of the first derivative of 1/$\xi(T)$. In Fig.\ref{figS7} we show the evolution of $T_0$ with $\Delta$. For the commensurate density, at $\Delta$=0, $k_BT_0$=0.23(6)$U$, in agreement with the predicted melting temperature for the Mott insulator $k_BT\simeq0.2U$ \cite{Gerbier}, while it decreases with increasing $\Delta$, consistently with a reduction of the gap due to the disorder. The BG does instead show a linear increase of the crossover temperature, $k_BT_0\propto\Delta$. This result, already observed in numerical simulations at small disorder strength \cite{Nessi}, can be intuitively justified with the $\Delta$-scaling of the effective Fermi energy of the BG. The different crossover temperature in the incommensurate and commensurate cases for small disorder suggests why the fit of the momentum distribution with only one $\xi_T$ is not working properly in this regime (Fig.4(c)). For larger $\Delta$, the simulations indicate that $T_0$ is comparable with the temperature measured experimentally in the SF regime, supporting the large $\xi_T$ observed in this regime.

\begin{figure}[ht!]
\includegraphics[width=0.95\columnwidth] {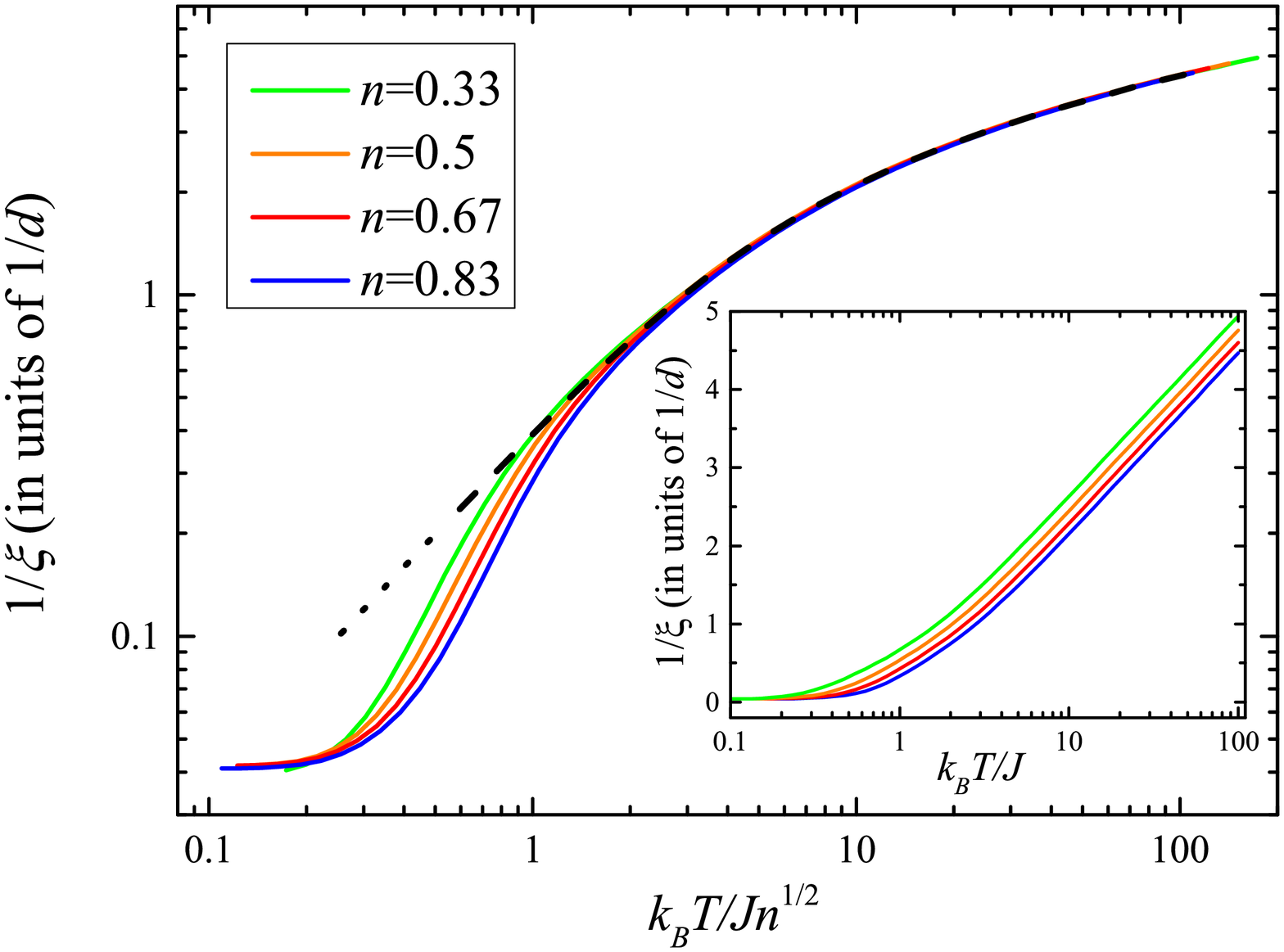}
\caption{SF correlation length vs density and temperature. Correlation length $\xi$ vs $T$, calculated by exact-diagonalization of a clean system with $L$=12$d$, for various site occupations. The dashed line is a fit of the high-$T$ data with eq.S1. Inset: density dependence.}
\label{figS8}
\end{figure}

\textbf{Temperature estimation in the superfluid regime}. An important application of the exact diagonalization study is the estimation of the experimental temperature from the measured correlation length in the SF regime ($\Delta$=0, $U\simeq J$). Finite-size effects prevent us from obtaining quantitative results for $\xi(T)$ at low temperature, but we are able to determine the correct high-$T$ scaling of the correlation length for the clean SF. Fig.\ref{figS8} shows more details of the derived scaling
$\xi$=$d/{\rm arcsinh}(k_B T/\alpha Jn^{1/2})$, which fits the numerical results in the regime $k_BT$=(2-100)$J$ and $n\leq$1 with $\alpha$=2.50(5). In this regime one can neglect the $T$=0 correlation length and identify $\xi$ with $\xi_T$. The results shown are for $U$=2$J$, but other calculations in the SF regime, suggest little or no dependence on $U$. To our knowledge, this behavior was not found in previous studies of bosonic systems. A similar result, $\xi\simeq d/{\rm arcsinh}(k_BT/J)$, was however found for spinless fermions in a lattice \cite{Sirker}, which represents the $U$=$\infty$ bosonic limit. The weak logarithmic temperature dependence for $k_BT>J$ can be attributed to the finite lattice bandwidth. For vanishing $T$, eq.S1 tends instead to the usual linear scaling in $T$ of the Luttinger liquid theory, $\xi_T\simeq d J/k_BT$ \cite{Giamarchibook}. The finite size of our simulations does not allow to study this regime, but the data in Fig.\ref{figS8} suggest that it is reached only for $k_BT\ll J$.

To adapt these homogeneous results to the trapped systems we have employed a local-density approximation, and averaged the finite-$T$ momentum distribution, $P(k)\simeq 1/(2\pi\xi_T(k^2+(1/\xi_T)^2)$, over the zero-$T$ density distribution in the trap. The average correlation length $\langle\xi_T(z)\rangle$ is very close to the one calculated for the peak density in the trap, $\langle\xi_T(z)\rangle\simeq0.90\xi_T(0)$, where the numerical prefactor has a very weak dependence on density and temperature in the experimental range. An additional average over the tubes distribution as in the experiment confirms that the average correlation length is $\langle\xi_T(z,i,j)\rangle\simeq 0.99\langle\xi_T(z,i_m,j_m)\rangle$, where $(i_m,j_m)$ represents a single tube with the peak density equal to the mean peak densities of all the tubes. This suggests that one can treat the collection of tubes as if they had the same thermal correlation length and formally justifies the assumption of a single thermal correlation for the SF regime.

\begin{figure}[ht!]
\includegraphics[width=0.95\columnwidth] {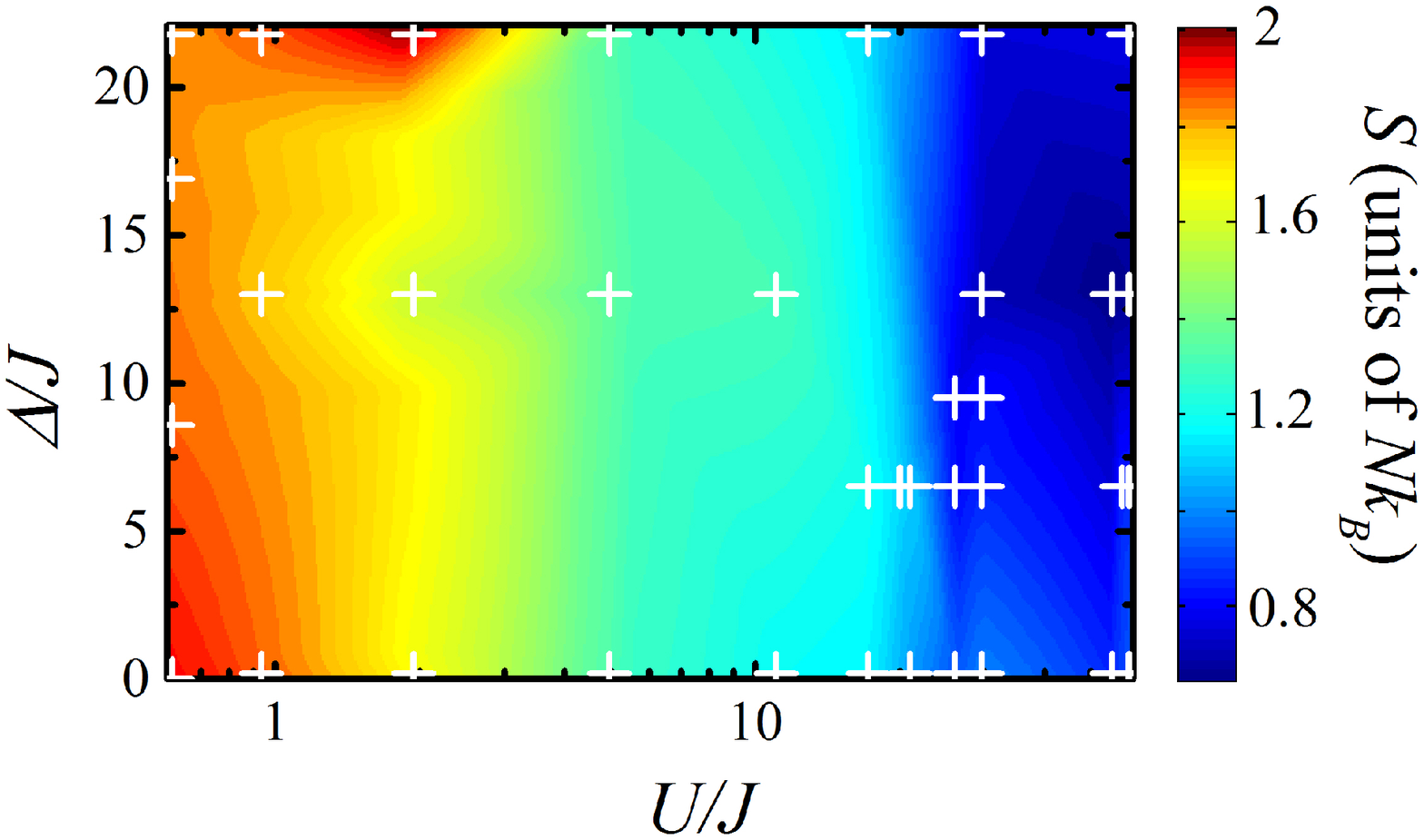}
\caption{Measured entropy per particle, $S/Nk_B$. The white crosses show the data points from which the 2D graph was created by interpolation.}
\label{figS9}
\end{figure}

In the measurements shown in Fig.1, for $U$=2.3$J$, from the fitted $\xi_T$=1.38(13)$d$, we obtain $k_BT$=3.1(4)(3)$J$. The first uncertainty is the statistical error on $\xi_T$ and $n$, and the second one is the systematic error on the calibration of $N_T$ ($\simeq$50\%). For lower interaction strengths we observe an increase of the temperature up to $k_BT$=5.4(8)(4)$J$ for $U$=0.4$J$. This suggests that the preparation of the system at small interactions is not fully adiabatic, as suggested also from the entropy measurement shown below (Fig.\ref{figS9}).

We observe a background heating of the SF component approximately linear with time, with a rate of 8.5(1.4)$J/s$ at $U$=$J$.  This heating is presumably due to phase and amplitude fluctuations in the lattices, and might justify the lowest temperature we can achieve in the experiment. We exploited this mechanism to realize measurements at higher temperature/entropy (grey squares in Fig.1(b) and inset of Fig.4(e)).

\textbf{Entropy estimation}. An extensive study of other regions of the $U-\Delta$ plane by exact diagonalization may in principle allow extracting a temperature from the experimental data also in presence of disorder, assuming thermal equilibrium. This is however a much more difficult problem, since in general there will be a coexistence of different phases, and even for a single phase one might expect a non-trivial dependence of the correlation length from all the parameters in the problem, $\xi_T$=$\xi_T (U, \Delta, n, T )$.

In the absence of a theory, we estimated a bound on the entropy of the system across the $U-\Delta$ plane. The protocol is the following: we measure the initial entropy of the system in the 3D trap; we measure again the entropy after having transferred the system into the 1D tubes and back into the 3D trap; we take the mean value of the initial and final entropies as an indication of the entropy in the 1D tubes. In the BEC regime of $T/T_c <1$, where $T_c$ is the critical temperature for condensation in 3D, we use the relation $S$=4$N_T k_B \zeta(4)/\zeta(3)(T/T_c)^3$, where $\zeta$  is the Riemann Zeta function \cite{Catani}. The reduced temperature $T/T_c$ is estimated from the measured condensed fraction by taking into account the finite interaction energy. In the thermal regime, $T>T_c$, we use the relation $S$=$N_T k_B [4-\log(N_T (\hbar\omega/k_B T)^3)]$.

The entropy measurements are shown in Fig.\ref{figS9}.  There is an overall increase of $S$ towards small $U$, which is presumably due to a reduced adiabaticity in the preparation of the 1D systems at small interactions. For this reason the temperature estimated in the small-$U$ region can be an over estimation of the temperature in regions with larger $U$.

\end{document}